\documentclass[10pt]{article}

\pdfoutput=1 

\usepackage{amsmath}
\usepackage{amsfonts}
\usepackage{amssymb}
\usepackage{textcomp}
\usepackage{hyperref}
  \hypersetup{
    colorlinks=true,
    urlcolor=blue,
    pdftitle={An example of geometric origami design with benefit of graph enumeration algorithms},
    pdfauthor={David Dureisseix}}
\usepackage{graphicx}

\title{An example of geometric origami design 
       with benefit of graph enumeration algorithms}
\author{David Dureisseix}
\date{}

\begin{document}

\maketitle
\begin{center}
  Univ Lyon, INSA Lyon (France)
\end{center}

\paragraph{Abstract:}
This article is concerned with an example of complex
planar geometry arising from flat origami challenges.
The complexity of solution algorithms is illustrated,
depending on the depth of the initial analysis of the problem,
starting from brute force enumeration, up to the equivalence to
a dedicated problem in graph theory.
This leads to algorithms starting from an untractable case on modern
computers, up to a run of few seconds on a portable personal computer.
This emphasizes the need for a prior analysis by humans before 
considering the assistance of computers for design problems.
The graph problem is an enumeration of spanning trees from a grid graph, 
leading to a coarse scale description of the geometry of the
paper edge on the flat-folded state.


\section{Introduction}

Origami (paperfolding without cutting nor gluing) and especially
flat-folded model design, strongly relies on planar geometry.
Color-changing technique, using appropriately a paper with
one color on a face and a second color on the other face,
adds some challenges to the previous model design.
The most demanding cases use numerous and alternated color changes
on the flat-folded state of the paper, and making a chessboard
(or checkered patterns) is among the hardest problems,
provided that one adds the constraint of starting from a single 
square sheet of paper. 
Indeed, starting from a narrow strip of two-colored 
paper, or from several separated sheets (a technique known as modular
origami), drastically reduces the difficulty of designing a
checkered pattern.

\begin{figure}
  \centering
  \includegraphics[scale=0.5]{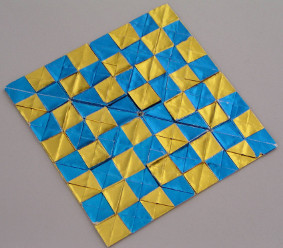}
  \caption{A possibly optimal $8 \times 8$ chessboards, from
  a $32 \times 32$ square of paper, after
  \cite{Dureisseix2000}.}
  \label{fig:chessboard}
\end{figure}

Several designs appeared during the past 30 years, 
see Table~\ref{tab:chessboard},
and the question of the optimality was settled:
to get a $n \times n$ checkered pattern, what is the
minimal size of the initial square sheet of paper?
In its present general form, this question is still open.
Nevertheless, once a design is completed, an upper bound for the
optimal initial square sheet size is made available.
Some estimates were also given:
with the sensible assumption that a color-change always appears 
along a side of the initial square sheet of paper, 
this leads to the question of the length of a continuous path
which is followed by the sides of the initial square sheet of paper
on the flat-folded state, called the edge diagram.
With such an assumption, the answer was that a half-perimeter
$s = n^2$ is required \cite{Dureisseix2000},
see figure~\ref{fig:chessboard}.
Note that this polynomial complexity is the one of a solution
(or the complexity of checking that a folded model is a solution),
but not the complexity of finding all the solutions
(i.e. of enumerating the solutions, not only counting them).

\begin{table}[htbp]
  \center
  \begin{tabular}{c c l}
    \hline
      Publication year & Initial paper size & \multicolumn{1}{c}{Author} \\
    \hline
      1985 & $64 \times 64$ & Hulme \cite{Hulme1985} \\
      1989 & $40 \times 40$ & Casey \cite{Casey1989} \\
      1993 & $36 \times 36$ & Montroll \cite{Montroll1993} \\
      1998 & $40 \times 40$ & Kirschenbaum \cite{Kirschenbaum1998} \\
      2000 & $32 \times 32$ & Dureisseix \cite{Dureisseix2000} \\
      2001 & $32 \times 32$ & Chen \cite{Chen2001} \\
      2007 & $32 \times 32$ & Hollebeke \cite{Hollebeke2007} \\
    \hline
  \end{tabular}
  \caption{Historical designs of some $8 \times 8$ chessboards,
  after \cite{Budai}.}
  \label{tab:chessboard}
\end{table}

\begin{figure}[htbp]
  \centering
  \includegraphics[scale=0.5]{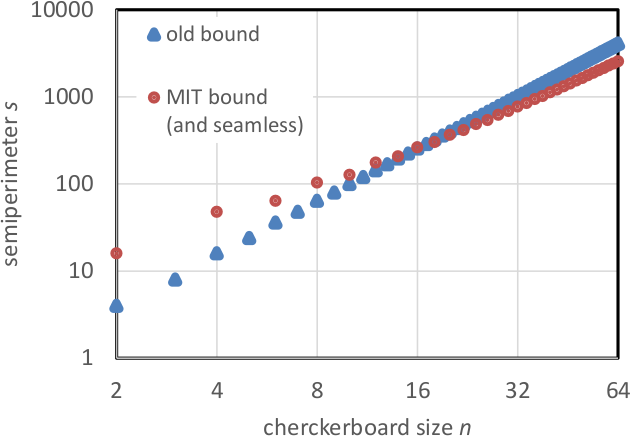}
  \caption{Bounds on complexity for the general problem of the 
           $n \times n$ checkered pattern.}
  \label{fig:MIT}
\end{figure}

2009 saw a breakthrough. 
Alleviating the previous constraint on the edge pattern, 
an asymptotic sharper bound has been given in \cite{Demaine2009} 
together with a general constructive proof. 
The authors found (for $n$ even):
$s = \frac{1}{2} n^2 + 8n + 8 - 5(n \mod 4)$. 
This bound is nevertheless outperforming the previous designs 
only for $n > 16$, see figure~\ref{fig:MIT}.
It also has the seamless property (each square board is made by a continuous
paper surface). 

Apart from the chessboard problem, another challenge emerged
recently on social networks: 
a design of a pixel matrix \cite{Tahira}. 
The difference with the pixel project \cite{childofsai2007} 
(for which small square modules of two possible different colors 
are assembled together to form a pixelated image,
the pixel unit of Max Hulme \cite{Hulme2012})
and the modular halftoning \cite{Xiao2015} 
lies on the use a single sheet of paper. 
Moreover, as for a LED matrix, each board square should be able to
change its color simply (e.g. with a single paper flip) 
and independently of the others. 
It appears that the proposed $8 \times 8$ pixel-matrix design can 
be obtained from a rectangular paper (an $8 \times 66$ strip is possible, 
a longer strip renders it easier; 
it is not known to the author if a shorter one is feasible). 
Therefore, the challenge of the design of an optimal $8 \times 8$
pixel matrix from a square sheet was settled.

This article focus on this last question, together with the 
possibility of using computers to check the possibilities and to help 
for designing, 
as well as the complexity of the associated design algorithms.

\section{Flipping mechanism and optimality challenge}

\begin{figure}
  \centering
  \includegraphics[scale=0.8]{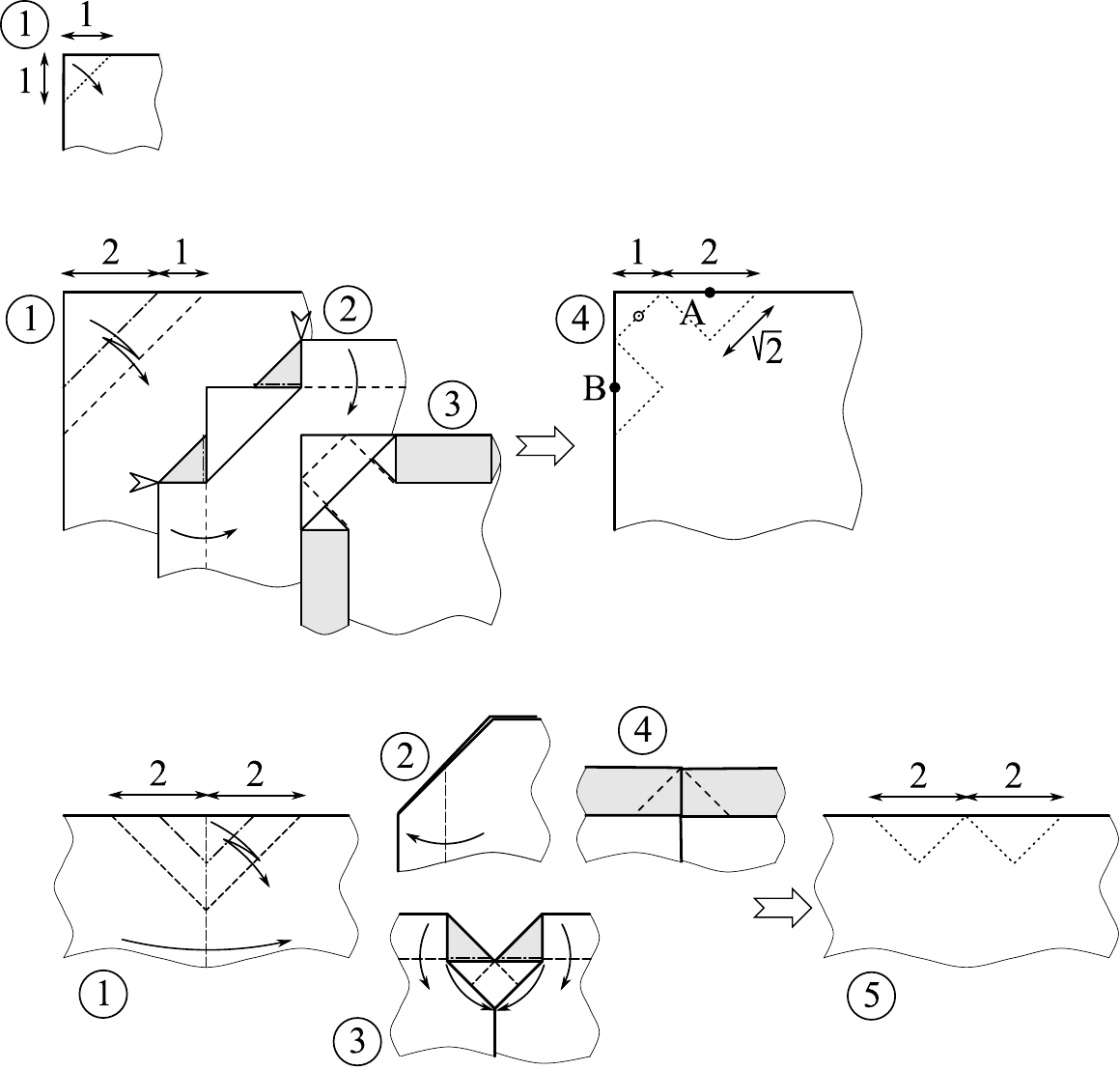}
  \caption{Folding diagrams of some flipping mechanisms. 
  Top: corner (simply flip the color of the corner);
  middle: corner + edges (1: crimp; 2: reverse; 
  3: flip the color of the board squares, unfold), 
  bottom: edges (1: crimp; 2: bookfold; 3: swivel folds; 
  4: flip, unfold).}
  \label{fig:flaps}
\end{figure}

The main argument used herein is to reuse the same edge assumption
as for the early chessboard: we still rely on the initial paper 
edge for the color-change, since it still produces the best known
paper optimality, at least for the $8 \times 8$ design.
Furthermore, a second assumption is needed for the flipping mechanism 
allowing a color-change on each board square independently
(and a simple single move);
we suggest to design an articulation on the diagonal of the board 
square. 
Using a corner or a side of the initial paper square therefore leads 
to two different folding designs; 
figure~\ref{fig:flaps} presents the elementary folding mechanisms to 
generate such flaps. 
The folded crease allowing each flap has a length of $\sqrt{2}$ 
and a perimeter length consumption of $2$, both for the corner and 
for the side mechanisms.

The former designs for classical static chessboard also possess some
diagonal flaps color-changing square board, but not for all of them.
The new design \cite{Demaine2009} also possess flaps; nevertheless,
they are square flaps along a board square edge and are not
independent on each board square (changing the overlap order would do
the job, but is not considered herein as a simple color-changing mechanism).

If this flipping mechanism is feasible, it will therefore 
require the use of $4$ corner mechanisms plus 
$n^2 - 4$ side mechanisms, leading to a semiperimeter with a minimum 
length of $n^2$, that is no more than the straight chessboard.
One could therefore challenge that an $8 \times 8$ color-changing
pixel-matrix can be designed with the same efficiency as for the 
best $8 \times 8$ chessboard, 
i.e. with a $32 \times 32$ square of two-colored paper.

In case of success, this would also exemplify the raise in difficulty 
by prescribing a design from a \emph{square} sheet of paper.
Indeed, the aforementioned strip ($8 \times 66$ rectangular paper)
leads to a mean thickness of the folded model (counted as the
average number of superimposed paper layers) to be
$t = (8 \times 66)/(8 \times 8) = 8.25$, while for the square paper,
it would raise for the hopefully best case to
$t = (32 \times 32)/(8 \times 8) = 16$.

\section{Design principles for the pixel matrix}

To design such a geometric origami model, straight force of
computers is not yet sufficient. 
Indeed, for flat-folding problems, the complexity of the task is 
very rapidly overwhelming.
An underlying basic question concerns the crease pattern,
which is the drawing on an unfolded flat sheet of paper of the
crease locations as well as their assignments (mountain or valley):
given a crease pattern, will it fold flat?
This question appeared to be not trivial. 
Some general conditions can be stated \cite{Justin1997} 
but they are hardly usable in practice; 
some necessary local constraints around each vertex 
(i.e. each crease intersection) 
\cite{Hull2002,Demaine2008,Abel2015,Evans2015} are nevertheless
easier to express, and one could expect relying on computers 
for running algorithms that could check the foldability.
Unfortunately, in their general forms,
these problems are hard to solve. 
For instance, the simple companion problem of layer ordering,
even with a given crease assignment, to decide if the model will
fold flat 
is NP-complete \cite{Bern1996}.
General case is therefore untractable, though some tools are already
available to help designing or checking rigid origami foldability,
such as \texttt{TreeMaker} \cite{Lang,Lang1996}, 
using circle packing \cite{Demaine2010}, 
and \texttt{Rigid Origami Simulator} \cite{Tachi,Tachi2009}, 
using mechanism theory.

Design searching therefore needs for intermediate steps in the
genuine problem, that may lead to more amenable solution strategies.
The proposal for the pixel-matrix design is to split the problem into
simpler subproblems: 
after the previous design for a flipping mechanism on the paper edge,
one could focus on an edge path determination, 
and on a fold propagation from the edge to the center of the paper.

\subsection{Generalized edge pattern and search complexity}

With the previous flipping mechanism, a first subproblem concerns
the placement on the flat-folded model 
(i.e. the $n \times n$ chessboard pattern)
of the articulation of the flap for each board square.
They should split each board square in half along one of its 
diagonal, and since they are connected to a paper initial edge
as in figure~\ref{fig:flaps},
the set of all those crease locations should
(i) be a continuous closed curve 
(it should possess only one closed loop), 
(ii) without crossing (but touching is allowed) 
since it should be on the top surface of the folded model, 
(iii) passing through one diagonal of each of the $n^2$ 
board squares.
This path could be named generalized edge pattern in reference to
its counterpart for the classical chessboard \cite{Dureisseix2000}.

As a guide for the design, the paperfolder is therefore interested
in selecting such a path that could be mapped onto the edge of the 
initial square sheet of paper. 
He could also select it with additional considerations such as 
symmetries (that may allow to reduced the size of the problem), 
estimates the difficulty of the task, etc. 
A useful information is therefore the enumeration of all 
possible paths.
This part of the problem is prone to computerization
and is discussed in the following.

\subsubsection{Brute force approach}

Without a deeper analysis, and since the path splits all the board squares 
in half, but with two diagonal possibilities each time, 
a crude enumerating approach consists in selecting one of the two possible 
diagonals per board square, for all the possible configurations, 
and testing for the aforementioned constraints.
This kind of problem is usually not polynomial in time, since the
difficulty lies in the number of possible selections of a set of 
$n^2$ diagonals: there are $p_1 = 2^{(n^2)}$ sets to test. 
This number increases rapidly with the size $n$; 
Table~\ref{tab:complexity} reports the corresponding values; 
the $8 \times 8$ case seems not to be possible to perform this 
way\footnote{The number of cases is close to the solution of 
the famous `wheat and chessboard' problem}.

\begin{table}[htbp]
  \small
  \center
  \begin{tabular}{c r r r r r}
    \hline
     $n$ & \multicolumn{1}{c}{$p_1$} & 
           \multicolumn{1}{c}{$p_2$} & 
           \multicolumn{1}{c}{$p_3$} &
           \multicolumn{1}{c}{$N$}   &
           \multicolumn{1}{c}{$\Tilde{N}$} \\
    \hline
 2 & 16      &  1 &  1 &  1 & 0 \\
 4 & 65\,536 & 16 &  4 &  4 & 0 \\
 6 & 68\,719\,476\,736 & 4\,096 & 495 & 192 & 11 \\
   & $\approx 69$ billions & & & & \\
 8 & 18\,446\,744\,073\,709\,551\,616 & 16\,777\,216 & 1\,307\,504 & 100\,352 & 3\,924 \\
   & $\approx 18$ billion billions & $\approx 17$ millions 
   & $\approx 1.3$ millions & & \\
10 &         &    &    &  557\,568\,000 &   \\
   &         &    &    & $\approx 558$ millions &   \\
    \hline
  \end{tabular}
  \caption{Number of cases to generate and test, 
  depending on the problem entry size $n$.}
  \label{tab:complexity}
\end{table}

\subsubsection{Path growing approach}

\begin{figure}
  \centering
  \includegraphics[scale=0.4]{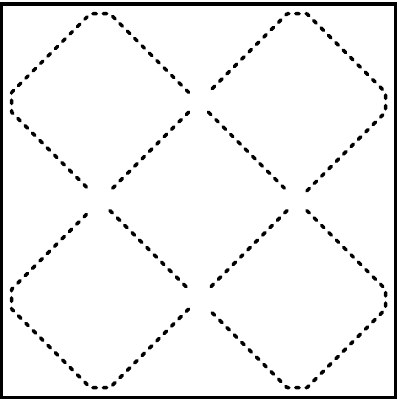}
  \includegraphics[scale=0.4]{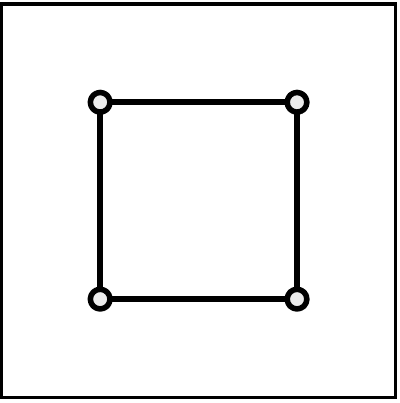}
  \includegraphics[scale=0.4]{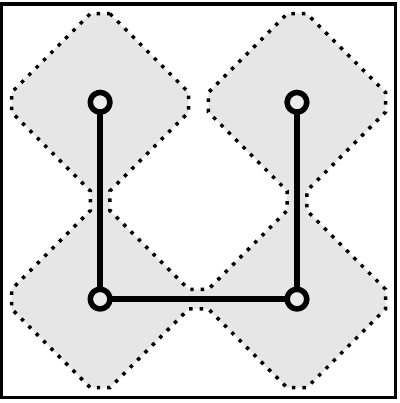}
  \caption{Path location for the generalized edge pattern of the 
  $4 \times 4$ pixel-matrix problem, and the corresponding graph. 
  The only solution, but which does not allow any corner placement 
  is also depicted.}
  \label{fig:44}
\end{figure}

\begin{figure}
  \centering
  \includegraphics[scale=0.4]{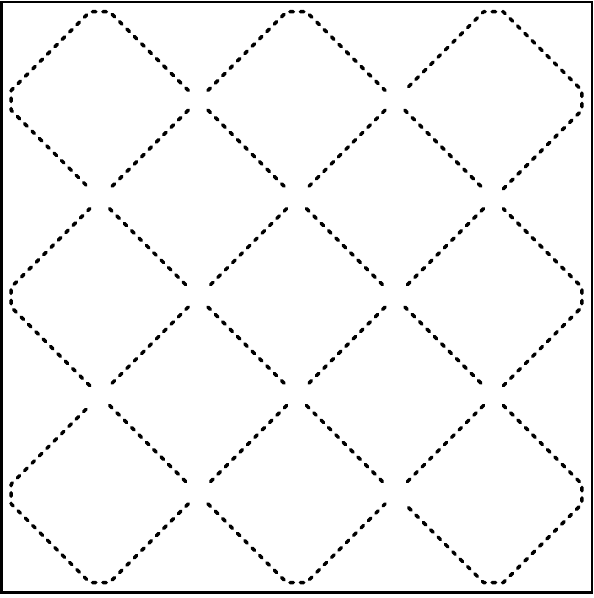}
  \includegraphics[scale=0.8]{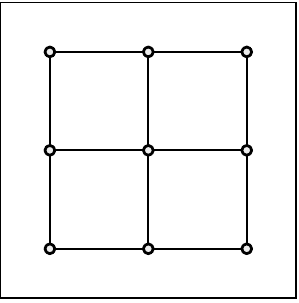}
  \includegraphics[scale=0.8]{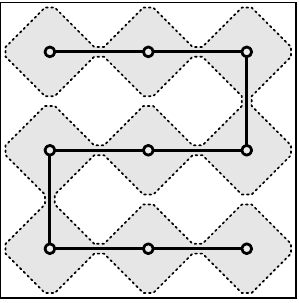}
  \caption{Path location for the generalized edge pattern of the 
  $6 \times 6$ pixel-matrix problem, and the corresponding graph. 
  One solution among others is also depicted.}
  \label{fig:66}
\end{figure}

\begin{figure}
  \centering
  \includegraphics[scale=0.8]{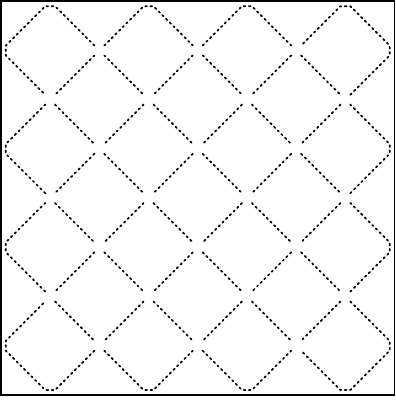}
  \includegraphics[scale=0.8]{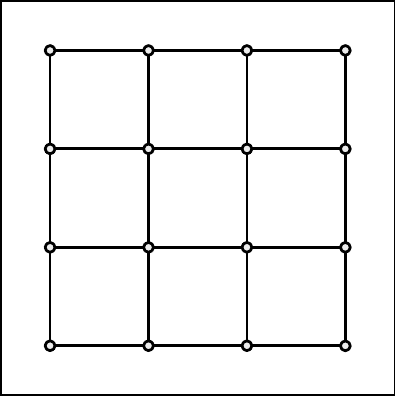}
  \includegraphics[scale=0.8]{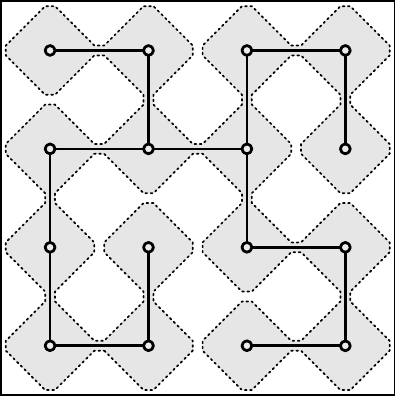}
  \caption{Path location for the generalized edge pattern of the 
  $8 \times 8$ pixel-matrix problem, and the corresponding graph. 
  One solution among others is also depicted.}
  \label{fig:88}
\end{figure}

\begin{figure}
  \centering
  \includegraphics[scale=1]{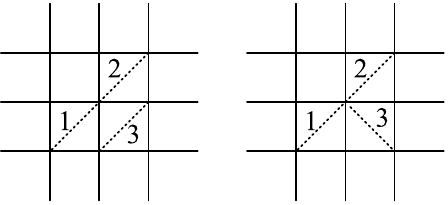}
  \caption{Local connections of the path when segments 1 and 2
  are aligned. 
  Left: segment 3 cannot be connected to the previous path of
  segments 1 and 2 due to its orientation; 
  right: segment 3 with the converse orientation has a pending end
  point and the path cannot be closed. 
  Conclusion: successive segments 1 and 2 cannot be aligned.}
  \label{fig:connexion}
\end{figure}

A second approach consists of making a non-crossing path grows 
in a continuous way, by step of one diagonal at a time 
(called a segment in the following) and 
exploring all alternatives, i.e. the 
possible different orientations for the next segment,
with backtracking.
Due to the closed and continuous characters of the path, 
not all the diagonal sites are feasible, 
opportunely reducing the problem size:
corner board squares could be split in only one way
(otherwise there is a pending segment at the corner, preventing the
path to be continuous), and for an even $n$ (only considered in the
following), the constraint propagates to allow only one feasible 
diagonal per board square.
The possible location for the path is depicted for a 
$4 \times 4$ problem in figure~\ref{fig:44} (left), for a
$6 \times 6$ problem in figure~\ref{fig:66} (left), and for an 
$8 \times 8$ problem in figure~\ref{fig:88} (left).  
Moreover all the successive segments are connected at right angle.
This could be understood as a consequence of the path requirements:
if two segments are aligned, the path should also contain 
a segment going perpendicular at the connection node 
(to fill all the board edges) see figure~\ref{fig:connexion}; 
this segment should have a pending end-point 
(it cannot cross the previous sub-path of the two segments);
therefore, the path could not be closed.

All in all, these criteria reduce the problem to the choice between 
two connections at each missing point, leading to a number of 
possibilities of
$p_2 = 2^{\left(\frac{1}{2}n^2 - n\right)}$. 
For each, the single-loop constraint should be tested. 
Though this number still grows rapidly with $n$, the 
$8 \times 8$ case is now possible to be computerized,
see Table~\ref{tab:complexity}.

The author programmed an algorithm to solve the path search problem
by making the path grows, which is a somehow engineering approach to 
find a way to get a solution.
From the $p_2 = 16\,777\,216$ cases, 
only $20\,826$ are single-loop non-crossing paths.
Among these last ones, some are symmetric to others, in
the symmetry group of the square (dihedral group of order 8). 
Once eliminating them, the number of solutions reduces
to $12\,600$.

\subsubsection{Pathway to the graph theory}
Due to the closed, continuous and non-crossing characters 
of the path, it splits the $n \times n$ square domain in 
an outer and an interior subdomains. 
These subdomains are composed by cells that are the union of 
half diagonal parts of the board squares sharing a corner
(called vertex in the following),
i.e. squares tilted by 45 degrees; 
figures~\ref{fig:44} and \ref{fig:88} (right)
depict these two subdomains in different gray levels.
Focusing on the interior gray subdomain, connected by cell corners
and whose contour is the searched path, an equivalent graph can be
defined. The graph vertices are the 
aforementioned $\nu = (n/2)^2$ previous vertices,
and the arcs relate all the vertices, without loop,
containing the connectivity information of the cells.
This is therefore a spanning tree of the square grid graph with $\nu$
vertices, known to have $\nu - 1$ arcs \cite{Tutte2001}.
The interest of this new problem formulation is to be able to rely
on numerous previous works on graphs.

The problem of finding all the possible paths is therefore casted
into a spanning tree enumeration.
$N$ denotes the number of these spanning trees, and is reported in 
Table~\ref{tab:complexity}. 
It can be obtained with the Kirchhoff's matrix-tree theorem 
\cite{Kirchhoff1847,Chaiken1978,Desjarlais2000}. 
For the $4 \times 4$ pixel matrix, one gets $N = 4$ 
(but all are symmetric of the first one --- mirroring or rotating --- 
so only 1 spanning tree is interesting, leading to a single solution for 
the path). 
This number also increases rapidly with the size of the board: 
for the $8 \times 8$ pixel matrix, one gets $N = 100\,352$. 
This is a particular integer sequence known as A007341 
\cite{Sloane1995,A007341}. 

A first approach lies in choosing the arcs of a possible spanning 
tree between those of the grid graph, and testing for each the nodes 
involved and the acyclicity. 
The number of graphs to be tested is then
$p_3 = C_e^{\nu - 1} = \frac{e!}{(e-\nu+1)!(\nu-1)!}$, 
where $e = n(n/2-1)$ is the number of arcs of the grid graph,
see Table~\ref{tab:complexity}. 

This approach is nevertheless not the most efficient. 
There are several available algorithms for enumerating all these 
trees, and their complexity are increasing as $O(N + \nu + e)$
(and $p_3$ is known to bound above $N$)\footnote{%
asymptotically, when $n$ is large, a closed-form expression for
$N$ is $e^{(\nu^2 \times 4C/\pi)}$
\cite{Shrock2000} where $C=1-1/3^3 + 1/5^2 - 1/7^2 \dots$ is
the Catalan constant, so that $N \approx 1.3385^{(n^2)}$
which is indeed an exponential growth,
to compare to $p_1 = 2^{(n^2)}$ and $p_2 \approx 1.4142^{(n^2)}$ 
}.
Few efficient stand-alone implementations have been made available, 
with the notable exception of the \texttt{grayspspan} code of 
D.~Knuth \cite{Knuth1993,SGB} that has been used herein. 
A graph contains essentially a topogical information, 
but with the node coordinates for the problem under concern, 
a geometrical information is available, and allows to detect the 
trees that are symmetric to an other one, as previously.
Once eliminating them, the number of solutions reduces from 
$N = 100\,352$ to $12\,600$
(indeed, the same as for the previous approach,
which is a good cross-check of the implementations).

Traducing the problem in term of graphs has  two main effects: 
first it restate the problem as a more generic one for which more 
efficient algorithms are available; 
second it reduces its size since the graph is built on a coarser grid 
than the path of the paper edge.
This second feature is kind of a multiscale problem as it can be seen 
on figure~\ref{fig:88} (right). 
The involved scales could be quantified by the lengths of the tree 
and path: for the spanning tree, the length of its $\nu-1$ edges is 
$2(\nu - 1)$ while the length of the generalized edge path is 
$\sqrt{2}n^2$. 
The length difference is related to the microscale so that the 
characteristic scale ratio is 
$(n^2-\sqrt{2}(\nu - 1))/n^2 \ge 1-\sqrt{2}/4 \approx 0.646$
(the asymptotic value for large $n$). 
The scales are therefore hardly separable.
It would certainly be interesting to be able to find even coarser 
models to reduce further the size of the problem, but such upscalings 
are somehow case dependent, and not easy to derive.
Once a macroscale spanning tree is found, the unique associated path 
should be built as a microscale corrugation. 

\subsection{Corner placement and contraction property} 

A classical necessary condition for crease patterns design
is a contraction property \cite{Dacorogna2008,Demaine2014}: 
the set of all fold intersection points
should contract from the unfolded stage to the folded one, i.e.
the distance between each pair of 
points should reduce (or be kept constant)
during the folding process.

\begin{figure}
  \centering
  \includegraphics[scale=1]{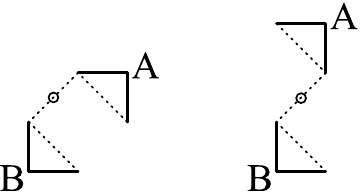}
  \caption{Generalized edge pattern around a corner. 
  Left: feasible path; right: unfeasible path.}
  \label{fig:corner}
\end{figure}

This should therefore apply on the boundary of the square
sheet of paper to the generalized folded path.
Since many pairs on points are involved
(with two possible positions for each flap, the number of distance
comparisons for $n$ even is $\frac{19}{2}n^4 - \frac{15}{2}n^2 - 2$), 
a simplified and weaker 
(though suspected to be the most constraint part of the problem)
necessary condition concerns first the 
corners of the initial square sheet of paper: 
while the generalized edge pattern of figure~\ref{fig:corner} (left) 
is feasible, this is not the case for the one of 
figure~\ref{fig:corner} (right).
This is a notable difference between the chessboard design 
\cite{Dureisseix2000} and the pixel-matrix design.
Indeed, the distance between points A and B on the flat-unfolded 
state figure~\ref{fig:flaps} (top, step 4) is 
$d = 2\sqrt{2} \approx 2.83$, 
while on the flat-folded state of figure~\ref{fig:corner} (right) 
it would be $d' = \sqrt{10} \approx 3.16$. 
However, the paper inextensibility requires 
a contraction by folding: $d \ge d'$,
and since it is not the case, the generalized edge pattern of 
figure~\ref{fig:corner} (right) is not feasible for a corner of the 
initial square of paper.

After building all the non-crossing one-loop continuous path of 
length $n^2$ from the spanning trees, one should check if they 
satisfy the corner placement constraint: 
there should be at least 4 segments equally spaced on the path whose 
predecessor and successor are not identically oriented (they may be 
aligned, but should not have the same orientation, 
figure~\ref{fig:corner}).
Among the feasible cases, the full contraction of the whole
edge of the initial square paper could then be checked.
 
For the $8 \times 8$ pixel matrix, 
among the previous $12\,600$ paths, this check should be made to eliminate
those that do not possess a feasible corner placement.
Doing so, only $\Tilde{N} = 3\,924$ paths remain, all
satisfying the contraction property 
(hence the strong constraint prescribed by the corner placement).
One can also note that for the $4 \times 4$ pixel matrix of
figure~\ref{fig:44}, 
the only feasible path does not have any possible corner placement; 
consequently the complexity for the semiperimeter is probably greater 
than $s = 4^2$ in this case
(maybe $s = 18$ is also not feasible, 
but $s = 20$ is somehow easy to fold).

\begin{figure}
  \centering
  \includegraphics[scale=0.6]{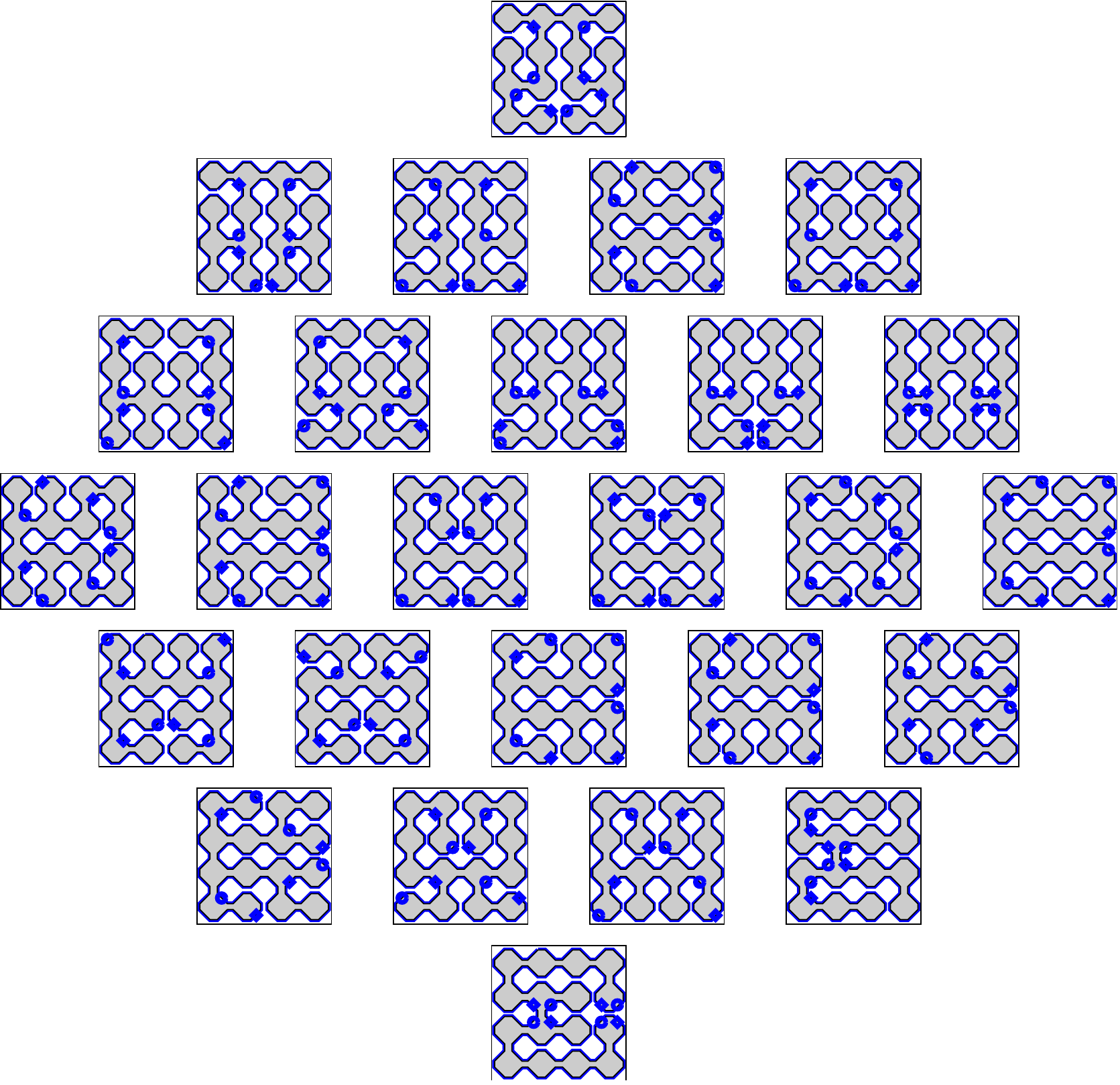}
  \caption{The $26$ self-symmetric paths and their corner placement
  possibilities (denoted by a set of 4 marks along the path).}
  \label{fig:26}
\end{figure}

\begin{figure}
  \centering
  \includegraphics[scale=0.6]{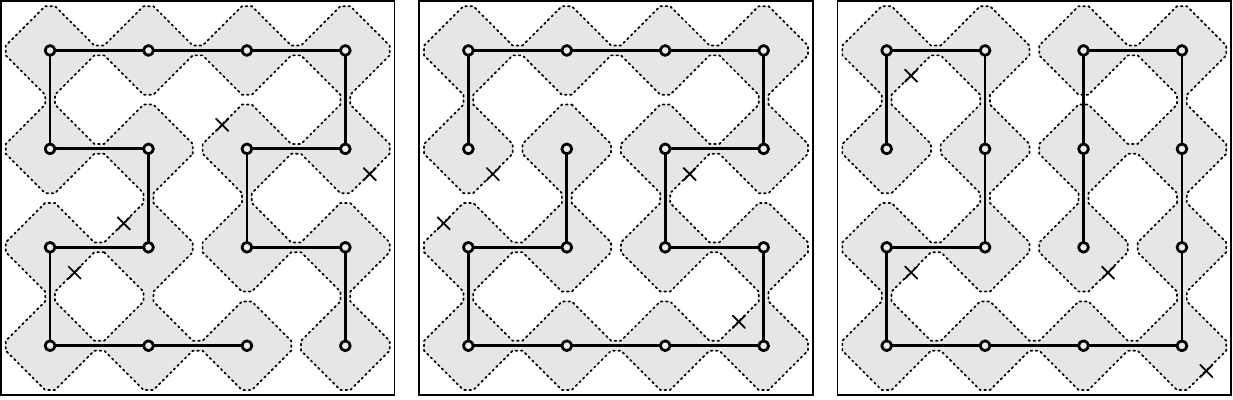}
  \caption{The $3$ paths of line spanning tree and their corner placement
  possibilities (denoted by a set of 4 marks along the path).}
  \label{fig:3lines}
\end{figure}

The interest in enumerating all paths with feasible corner placement
is to be able to select the a priori simpler or more suited cases.
For instance self-symmetry could simplify the design search problem.
For the $8 \times 8$ pixel matrix, among the previous $\Tilde{N} = 3\,924$ paths,
$26$ possess a vertical or horizontal self-symmetry
(none have any other self-symmetry);
they are depicted in figure~\ref{fig:26}, together with the different possible
corner placements. It is interesting to note that there are each
time 2 possible corner placements, each symmetric to the other,
but that none have a single self-symmetric corner placement.
As a consequence, and contrary to the straight chessboard case, 
the problem cannot be reduced by symmetry to part of the board.
A second attempt for a potential simplification lies in a spanning tree which is a
single line, therefore with only two end points (or leaves).
Only $3$ solutions with a feasible corner placement exist, all being contractive;
they are depicted in figure~\ref{fig:3lines}.

\subsection{Compatible fold propagation} 

Once a generalized edge pattern has been selected,
the last part of the problem is to fold flat a square sheet of paper
that maps its edges (and corners) on the pattern.
This part is more difficult to formulate in a way that is easily
computerizable.
Nevertheless, there have been at least two proposals that could lead
to a computer help program for this task, up to the author knowledge. 

\begin{figure}[htbp]
  \centering
  \includegraphics[scale=0.5]{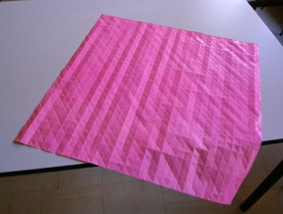}
  \includegraphics[scale=0.4]{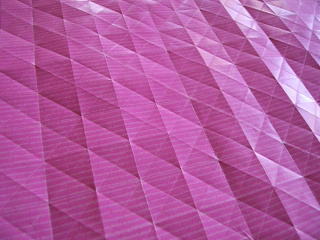}
  \caption{Pre-creasing a $32 \times 32$ square paper 
           to make a lattice of creases.}
  \label{fig:prepare}
\end{figure}

The first one is a systematic search on what could be called a `lattice origami' 
pattern \cite{Konjevod2009}. 
It assumes all the creases lying on a regular pattern on the initial square sheet 
of paper, consisting of vertical and horizontal creases distant of a unit value, 
and on the $\pm 45$\textdegree{} creases diagonalizing all the previous grid 
squares. 
They are depicted in figure~\ref{fig:prepare}. 
Such a crease lattice is kind of a discretization of the possible crease pattern 
family and so, it reduces the problem size and makes it more suitable for a 
discrete treatment by a computer. 
Unfortunately, this problem is still too computational demanding to be solved by 
brute force computing.

The second recent publication on this topic concerns an advance in solving
the problem of filling a hole in a crease pattern \cite{Demaine2014}. 
One of its instance consists in finding a flat-folded state for an initial
polygonal sheet of paper, allowing its boundary to match a prescribed 
path. This is in practice always possible, and it has been shown that an
algorithm can solve this problem in a polynomial time.
The only restriction lies in the fact that paper non self-intersecting is
not part of the constraints, but it render one confident in the fact that
this problem could be solved in reasonable time, though the general algorithm
is still an open question.

Up to now, with the previously mentioned restrictions,
the problem of fold propagation from the generalized edge pattern 
still relies on the skills and intuition of the paperfolder.
Several helping sketches are proposed in the following.

\subsubsection{A 2-scale approach?}

With the previous graph description of the problem, the multiscale
feature could hopefully be used to reduce the difficulty of the 
current task.
Indeed, one could proceed in two steps:
\begin{itemize}
  \item the fine scale corrugation, 
  using the proposed flipping mechanisms,
  could be folded on the border of the initial sheet of paper,
  reducing its size to a smaller flat-folded model.
  \item this last model could be considered as a sheet of its own,
  and its smaller perimeter could be mapped on the coarse scale
  corrugation, i.e. the graph path.
\end{itemize}
If feasible, this approach has the advantage of separating the scales in
this second part of the folding problem.
The edge mapping is still an issue, but has to be performed on a 
reduced-size problem (the coarse problem only).

The main issue relies on the scale separation.
Indeed, dealing with discrete geometry,
the microscale corrugation constrains the size of the
coarse sheet of paper. 
A side of the flat-folded coarse sheet should provide 
two half-corner mechanisms and a particular number, say $m$, 
of pairs of edge mechanisms.
The corresponding side length of the unfolded sheet is therefore
$a = 2 \times 3 + 4m$.
For a square paper, this length should equals half of the optimal 
semiperimeter, so $a = n^2/2$.
With $n$ even, one gets $2m = n^2/4 - 3$.
Nevertheless there is still an issue when starting
from a square sheet of paper:
since $m$ should be an integer, $n$ should not be a multiple of 4,
i.e. $n=4k+2$.
In this case, $m = 2k(k+1)-1$. 
This is feasible for $n=6$ but not for $n=8$.
When $n$ is a multiple of 4, 
this won't apply, though an almost-square solution is possible:
searching for a rectangular sheet of size $a_1 \times a_2$
with $a_i = 2 \times 3 + 4m_i$, $a_1+a_2=n^2$ and say, $n=4k$,
one gets $m_1+m_2 = 4k^2-3$ for which a solution is
$m_1 = 2k^2-2$, $m_2=2k^2-1$ and
$a_1=n^2/2-2$, $a_2=n^2/2+2$.
Therefore, for $n$ being a multiple of 4,
the scales are somehow entangled, and prevent the previous solution
procedure for a square coarse folded paper.
Note also that the almost-square case is always slightly less complex 
that the square case since the mean thickness is 
$t = a_1 a_2 / n^2 = n^2/4 - 4/n^2$
rather than 
$t = a^2/n^2 = n^2/4$.

Next section proposes a more direct solution to attempt the design
from a square sheet of paper, whose side can be a multiple of 4,
as for the $8 \times 8$ pixel matrix.

\subsubsection{Onion layers strategy for compatible fold propagation} 

Going back to the initial edge mapping,
the problem is at least two-fold:
(i) the edge of the square paper should map the selected pattern, and
(ii) the tortuosity of this path leads to a 2-scale corrugation on 
the edge of the paper, that have to be propagated from the edge up 
to the center of the paper, while keeping the folded state flat.

\begin{figure}[htbp]
  \centering
  \includegraphics[scale=0.4]{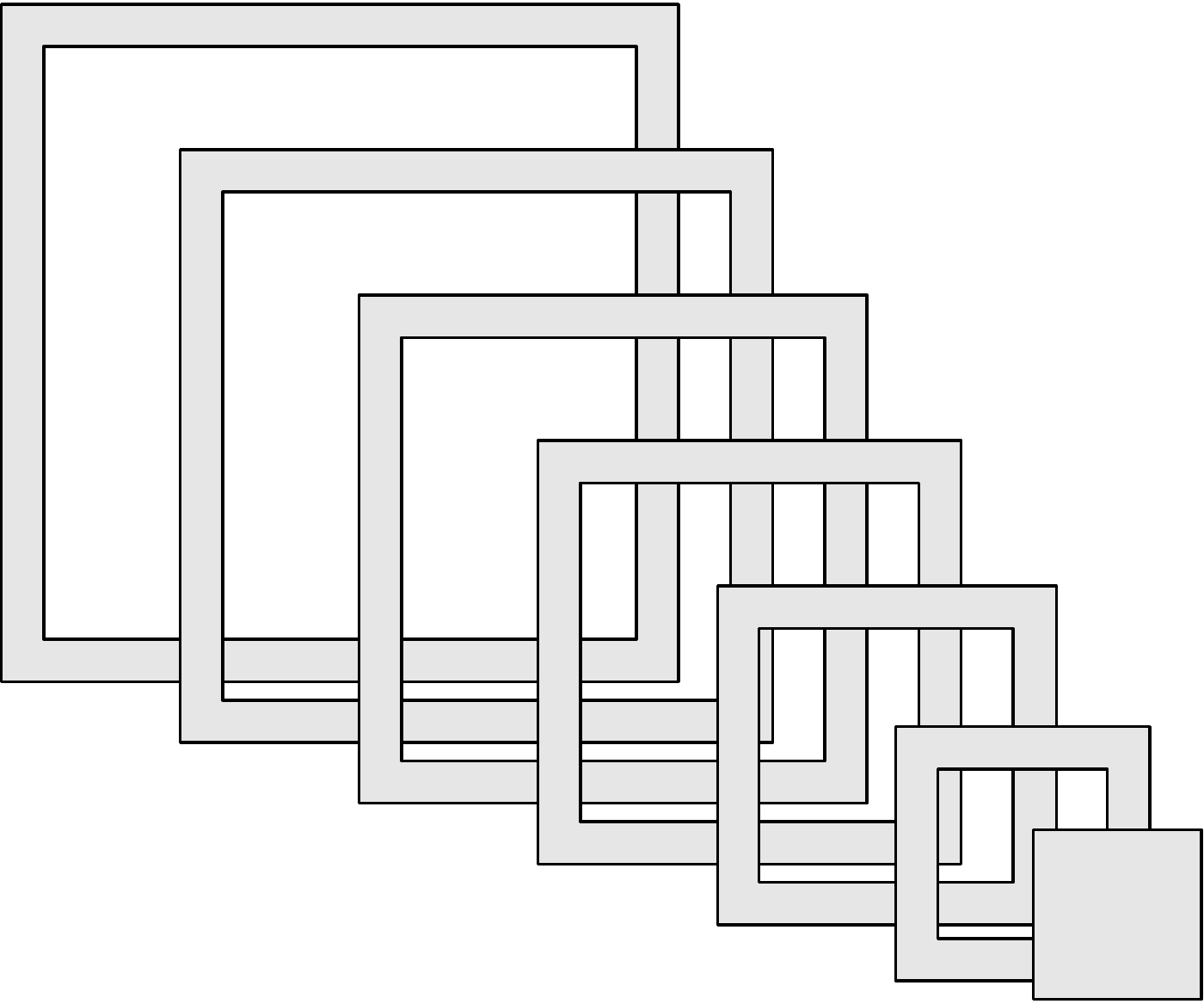}
  \includegraphics[scale=0.4]{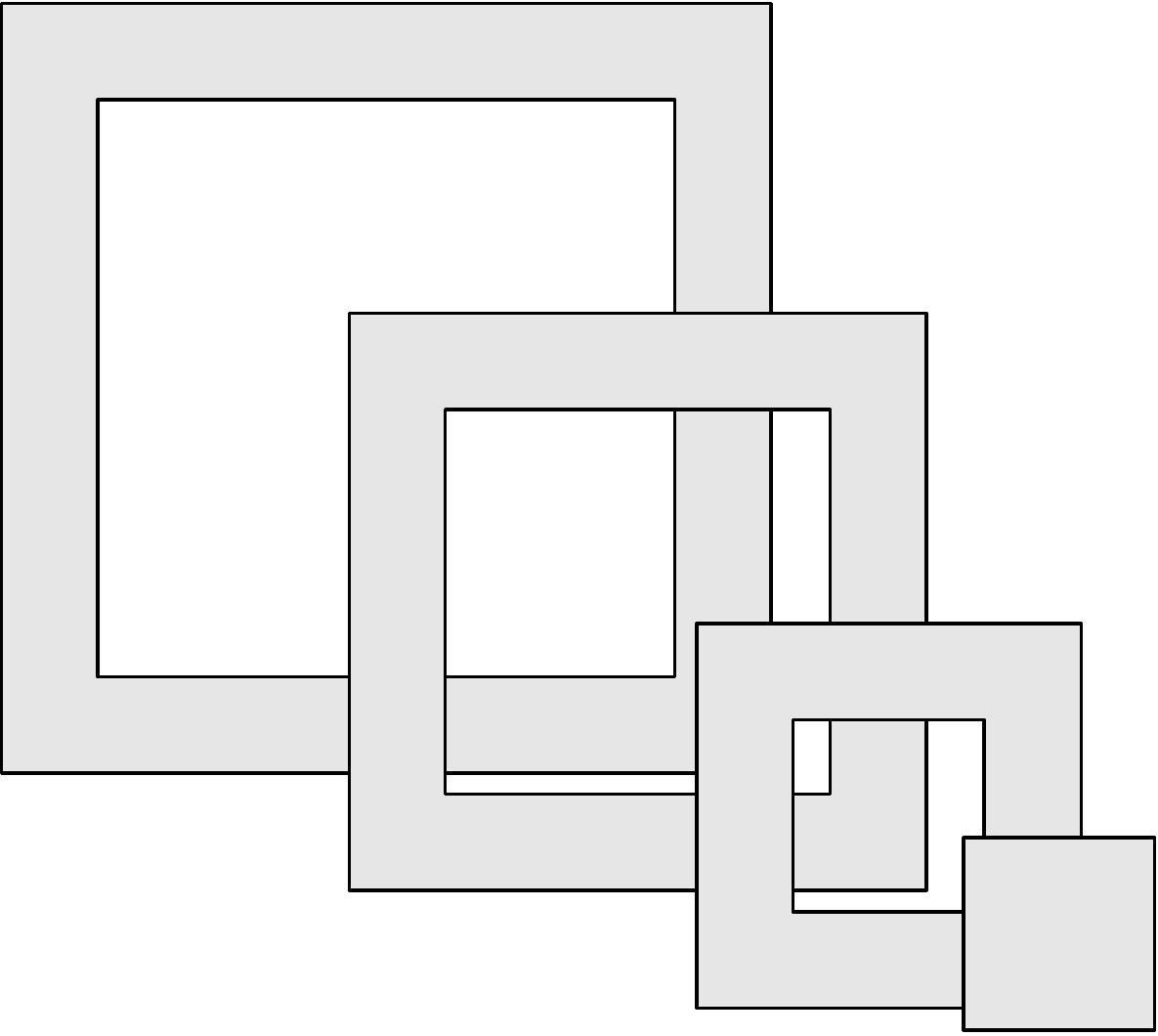}
  \caption{Successive sub-problems with decreasing perimeters, 
  for two choices of width.}
  \label{fig:onion}
\end{figure}

In addition to using the lattice creases, we now proceed by 
dealing with successive sub-problems.
We propose to cut the paper square in several parts that will have 
to be re-assembled at the end. 
If successful, this will lead to a prototype for the model design. 
The parts are herein successive nested rings lying on perimeters of 
decreasing sizes, figure~\ref{fig:onion}.
The successive problems consist in flat-folding a part, 
while mapping its perimeter to the internal side of the 
previously assembled part. 
Since the perimeter is decreasing, the size of the problem reduces,
but a folding compatibility should be satisfied to propagate the 
folds to the next part to be assembled. 
If and when the last central square is successfully assembled, 
the prototype is complete.

Selection of the width of the strips is an interesting issue. 
The goal is two-fold: 
each strip should map on the previous edge pattern, and 
flat-foldability has to be ensured. 
A thin strip would renders its flat-foldability easy, 
but do not solve enough foldability compatibility constraints, 
propagating too much the difficulty to the next inner strip, 
concentrating progressively the difficulty. 
Using a large strip allows to solve more progressively this issue 
but is more difficult to map on the required edge pattern. 
There is therefore a compromise to be found by choosing accordingly 
the onion layer width. 
For the $8 \times 8$ pixel matrix problem, 
a suited choice seems to be the one of 
figure~\ref{fig:onion} (right), with a width of 4.
The problem is solved once the central $8 \times 8$ square 
has been successfully mapped to the last perimeter.

\begin{figure}[htbp]
  \centering
  \includegraphics[scale=0.6]{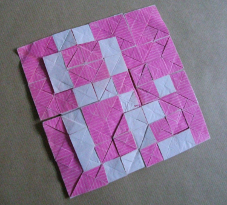}
  \caption{An example of a complete prototype.}
  \label{fig:model}
\end{figure}

A successful prototype is reproduced in figure~\ref{fig:model},
allowing to conclude that the $8 \times 8$ pixel matrix has
the same bound on complexity as the chessboard.

\section{Conclusions}

This article sets the pixel-matrix challenge for the point of view
of folding optimality, i.e. constraining the design of the given
origami model by prescribing an initial paper sheet to be square.
This appears to be a strong constraint that drives the complexity
of the task.
For a checkered $n \times n$ pattern design without transformable 
color-change (a static pattern), a bound on the semi-perimeter of 
the initial square of paper was $s = n^2$, 
having been improved only for $n$ larger than 16.
The pixel matrix adds the complex feature of a flipping mechanism 
allowing the local color-change independently on each board square.
Though adding a somehow significant difficulty on the design
(traduced here by a complex corner placement),
a surprising result leads to the same complexity for the
$8 \times 8$ design, as for the chessboard.

Concerning the design process, 
it has been exemplified that the brute force of computers is still 
unable to tackle problems we wish to solve in flat geometrical 
origami. 
The main issue remains the ability for the user to express 
the considered problem in a suited form, and especially to split it
in several well-chosen consecutive sub-problems, some of them being 
susceptible to be computerized. 
In any case, a deeper analysis is required to reduce the problem 
size, and express the problem within known scientific fields 
(such as graph theory).
For the current example, several analysis with increasing depths
lead to a succession of problem size reductions. 
Then the computer can be helpful to provide information on the 
potential designs. 
Nevertheless, several steps still hold on intuition and skills of 
human paperfolder to complete the design. 
In this sense, the computer may help, but cannot be substituted to 
the human for this task. 
This example illustrates the fact that the dramatic announcement 
of predominance of computer science on employability,
when commenting the publication \cite{Frey2013}
(there are also other predictions with significantly different 
fractions of automatizable jobs, for instance \cite{COE2017}),
could certainly emphasizes the complementary and gain 
to derive human-computer cooperation, 
though computers will lack to be autonomous in solving problems 
that are ill-posed. 
The human-computer interface will probably be a key issue, 
and problems need to be translated in computer tackling world;
in such a way we indeed need to transform our way of working
\cite{Brynjolfsson2014,Shanahan2015}. 
Such issues are also discussed about computer science education and 
learning programs, pros and cons are discussed and currently debated 
on the utility of early coding courses and/or on recasting 
traditional courses with a numeric culture orientation, as well as
on addressing the issue of the tutor education. 

\bibliographystyle{plainurl}
\bibliography{pap}

\clearpage
\appendix
\section{Some pixel matrix designs}

For a $n \times n$ static checkerboard, the optimality argument
(satisfied for $n < 16$) states that the smallest square
of paper has a semiperimeter $s = n^2$ for an even $n$.
The following examples illustrate the feasibility of the
same bound for the pixel matrix.

\subsection{The $2 \times 2$ pixel matrix}

The $2 \times 2$ pixel matrix is obtained with a single design from a 
$2 \times 2$ square of paper, following figure~\ref{fig:22design}.
It is nevertheless somehow disappointing since the lack of wasted paper
does not allow to provide additional underlying paper when folding the corners.
This is an edge effect that can be discarded but with the price to use
a sub-optimal $4 \times 4$ paper (whose design is left to the reader).

A $2 \times 2$ checkered board in \cite{Demaine2009} uses a
$3 \times 3$ paper, but cannot flip its color easily.
The design of appendix~\ref{sec:s222} can, but still have
an edge effect defect of 1 board square.
Therefore, it is unlikely that a design without such edge
effect can be folded with a square paper smaller than
$4 \times 4$.

\begin{figure}[htbp]
  \centering
  \includegraphics[scale=1.2]{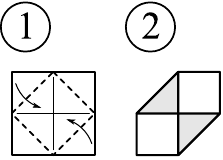}
  \caption{The optimal $2 \times 2$ pixel matrix, from
  a $2 \times 2$ square of paper.}
  \label{fig:22design}
\end{figure}

\subsection{$4 \times 4$ pixel matrix}

\begin{figure}[htbp]
  \centering
  \includegraphics[scale=1]{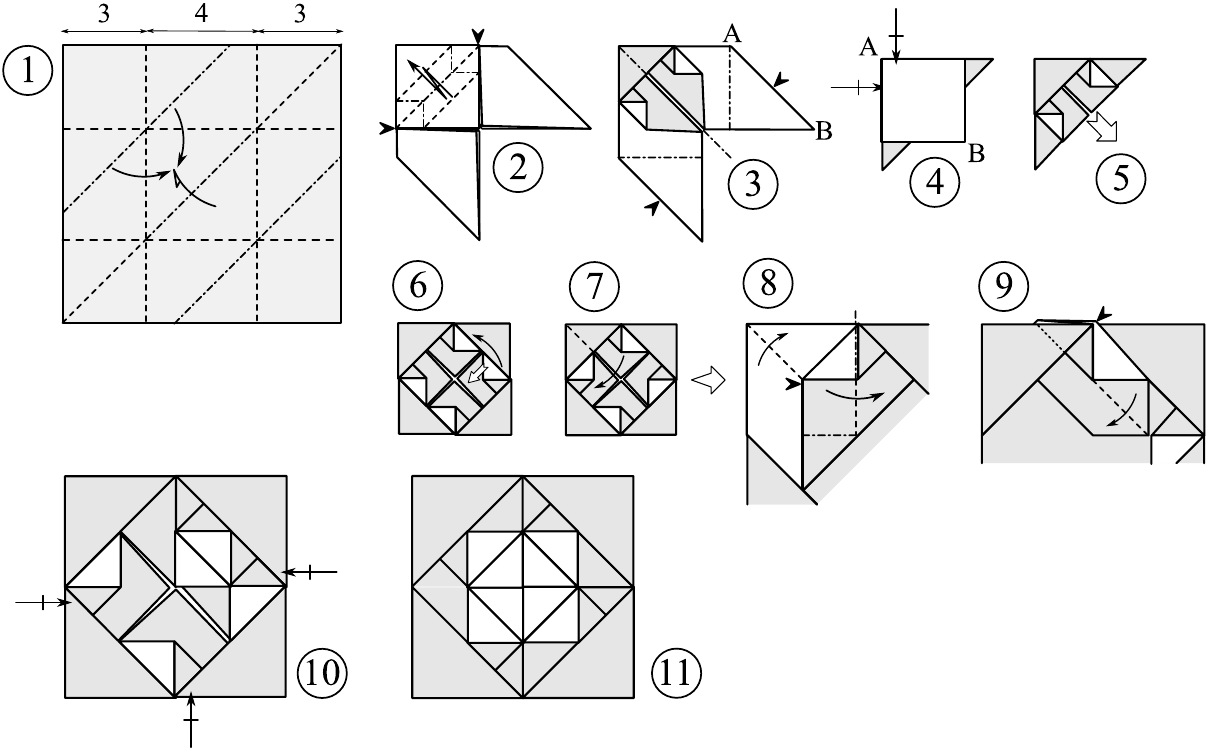}
\begin{itemize}
\item Step 1: from a $10 \times 10$ grid, collapse the `preliminary-like' folds.
\item Step 2: repeat backwards.
\item Step 3: open and refold on the second diagonal.
\item Step 4: repeat step 2 on both faces.
\item Step 5: half open.
\item Step 6: render the model central-symmetric.
\item Step 7: book fold.
\item Step 8: reverse and book fold back.
\item Step 9: reverse and sink.
\item Step 10: repeat three times the sequence of steps 7-9.
\end{itemize}
  \caption{The possibly suboptimal $4 \times 4$ pixel matrix, from
  a $10 \times 10$ square of paper.}
  \label{fig:44design}
\end{figure}

The $4 \times 4$ pixel matrix cannot be obtained with a square whose 
semiperimenter is $s = 4^2 = 16$,
i.e. with a $8 \times 8$ square paper.
There are several possible designs from a $10 \times 10$ paper;
one is depicted in figure~\ref{fig:44design}.

\clearpage
\subsection{$6 \times 6$ pixel matrix}

There are solutions for a $6 \times 6$ pixel matrix 
form a seminperimeter $s = 6^2 = 36$, i.e. with a 
$18 \times 18$ square of paper.
A design with a central symmetry is depicted in 
figure~\ref{fig:66} and is a solution of the 
previous sub-problem of finding a path in a coarse graph.
This step can be seen as a macroscopic problem
while the detailed flipping mechanism folding for color-changing
is a microscopic corrugation along the paper edge.

\begin{figure}[htbp]
  \centering
  \includegraphics[scale=0.8,angle=90]{solution66pixel}
  \includegraphics[scale=0.8,angle=90]{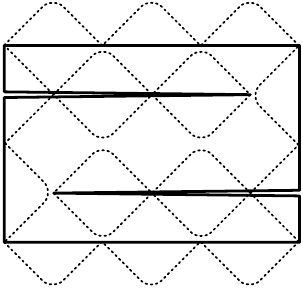}
  \caption{A path location for the generalized edge pattern of the 
  $6 \times 6$ pixel-matrix problem, and the corresponding 
  coarse edge location of length $4c=40$ supporting the fine corrugation.}
  \label{fig:solutionbord66pixel}
\end{figure}

Finding the complete folding sequence can then be studied
by at least two means:
\begin{itemize}
  \item propagating the folds from the edge once mapped on
  the targeted final folding state (the edge diagram);
  \item using a two-scale approach.
\end{itemize}
This last case may well be available, at least for even $n$ 
of the form $n=2(2k-1)$ with an integer $k$
(so that $n=2,6,10,14\ldots$). 
Indeed, for this case, the flipping mechanisms can be
folded flap to a `coarse' square configuration of
edge length $c=(n/2)^2+1$.
Then one can search for designing a folding sequence 
of this $c \times c$ square as a whole, mapping its
`coarse' perimeter on a dedicated edge pattern built
from the previous spanning tree (figure~\ref{fig:solutionbord66pixel}).
If this pseudo $c \times c$ square can be folded this way,
then the problem is fully separated on the two scales. 
Whether this is a general property for any $n$ or not
is still an open question.

For the $k=2,n=6,s/2=18,c=10$ case,
the `coarse' problem is not too complex, using the central
symmetry of the design, but cose not fully statisfy to the
lattice origami principle.
Indeed one issue is a shift of value 2 in 
figure~\ref{fig:solutionbord66pixel} (step 3)
leading to a `shearing' of the square.
A twist fold can be made as for some tesselation designs
\cite{Gjerde,Hull2013}, than involves an angle $\alpha$ with
$\tan \alpha = 2/6 = 1/3$  that cannot lie on the
($0$, $45$\textdegree, $90$\textdegree) $c \times c$  regular grid. 

\begin{figure}[htbp]
  \centering
  \includegraphics[scale=1]{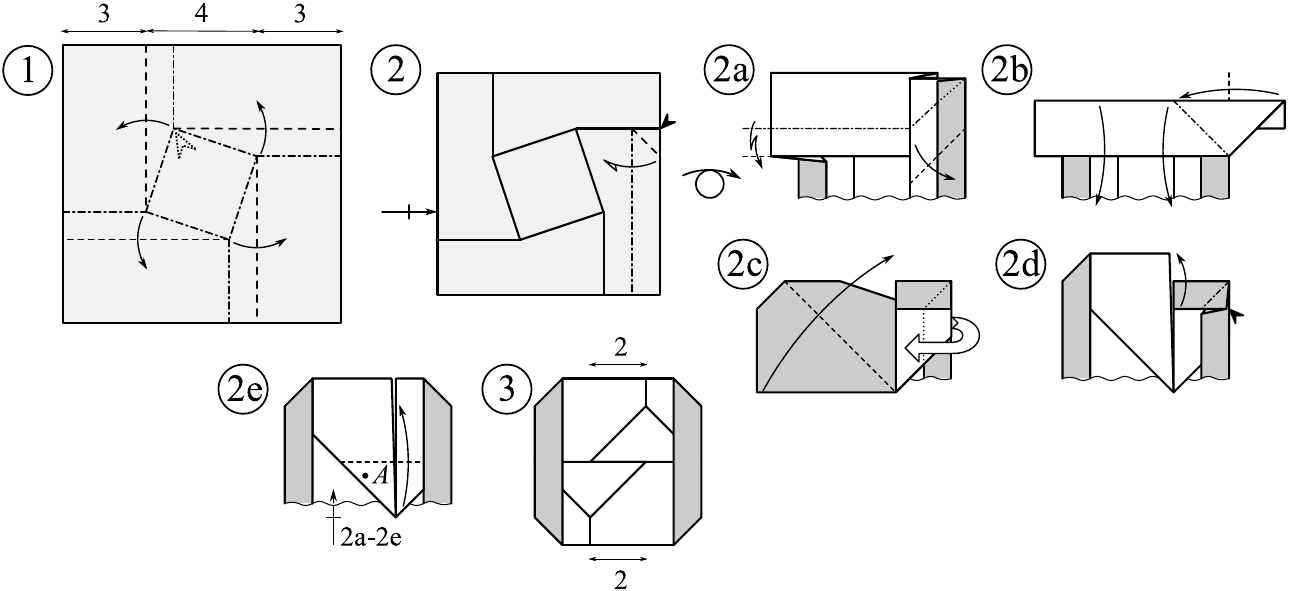}
\begin{itemize}
\item Step 1: twist fold.
\item Step 2: fold behind. Turn the model.
\item Step 2a: complex pleat, pivoting right part.
\item Step 2b: unfold and swivel fold.
\item Step 2c: book fold. Rever inside flap outside.
\item Step 2d: squash fold.
\item Step 2e: fold the flap to avoid subsequent interference.
  Repeat previous steps at the bottom.
\end{itemize}
  \caption{The coarse folding of a $10 \times 10$ square to
  a $6 \times 6$ pattern intended to be the coarse pixel-matrix.}
  \label{fig:model66pixelcoarse}
\end{figure}

Figure~\ref{fig:model66pixelcoarse} provides a folding sequence.
Note that for this coarse model, there is some paper self-intersection
in the vicinity of point $A$ at step 2e,
when repeating the sequence on the half-bottom part
(to avoid this, the book fold of step 2e is mandatory).
This cannot be solved at the only coarse scale, and is postponed to the
complete pixel-matrix folding sequence,
leading to conclude that the two scales may well be not
fully separable.

\clearpage

\clearpage
\section{Some static checkered boards}

Note for completion, that there is a series of origami puzzles
based on checkered patterns, willing to obtain the smallest number
of folds \cite{Grabarchuk,Mitchell2011} 
whereas we are herein interested in the smaller waste of paper.
For a $n \times n$ static checkerboard, the optimality argument
(satisfied for $n < 16$) states that the smallest square
of paper has a semiperimeter $s = n^2$ for an even $n$,
and $s = n^2 - 1$ for an odd $n$.
Some designs are given in the following sections.
One can note that there is a constructive proof of the previous bounds
for $s$; indeed, a recursive construction is given in the last following
sections.

\subsection{$2 \times 2$ checkered board}
\label{sec:s222}

\begin{figure}[htbp]
  \centering
  \includegraphics[scale=1]{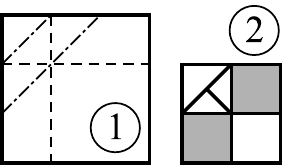}
  \caption{The suboptimal $2 \times 2$ checkerboard, from
  a $3 \times 3$ square of paper.}
  \label{fig:s222}
\end{figure}

As previously mentioned, the $2 \times 2$ case can be folded from
a sub-optimal $3 \times 3$ paper  
\cite{Dureisseix2000}, 
figure~\ref{fig:s222}.

\subsection{$3 \times 3$ checkered board}
 
\begin{figure}[htbp]
  \centering
  \includegraphics[scale=1]{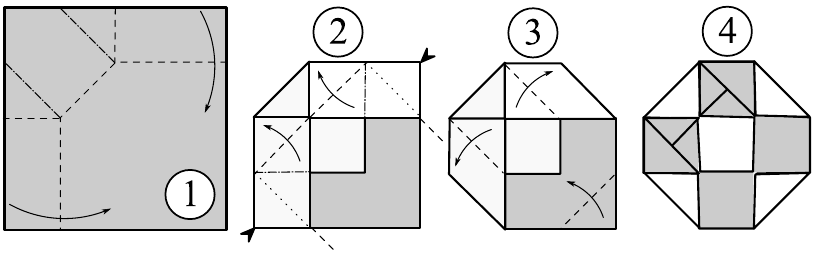}
\begin{itemize}
\item Step 1: from a $4 \times 4$ grid, fold along the crease pattern.
\item Step 2: two swivel folds.
\item Step 3: color change.
\end{itemize}
  \caption{The optimal $3 \times 3$ static checkerboard, from
  a $4 \times 4$ square of paper.}
  \label{fig:s33}
\end{figure}

Following the technique of the edge diagram,
the $3 \times 3$ checkered board requires a semiperimeter of
$8$, so a $4 \times 4$ square sheet of paper.
There are 3 possible corner placements, one of them
leading to the folding sequence of
figure~\ref{fig:s33}.

\begin{figure}[htbp]
  \centering
  \includegraphics[scale=1]{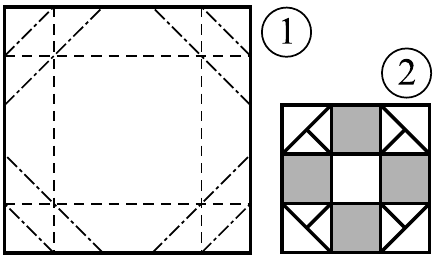}
  \caption{The sub-optimal $3 \times 3$ static checkerboard, from
  a $5 \times 5$ square of paper.}
  \label{fig:s332}
\end{figure}

It is also disappointing since the lack of wasted paper
does not allow to provide additional underlying paper when folding the corners.
This is an edge effect that can be discarded but with the price to use
a sub-optimal $5 \times 5$ paper \cite{Dureisseix2000}, 
figure~\ref{fig:s332}.

\subsection{$4 \times 4$ checkered board}

\begin{figure}[htbp]
  \centering
  \includegraphics[scale=1]{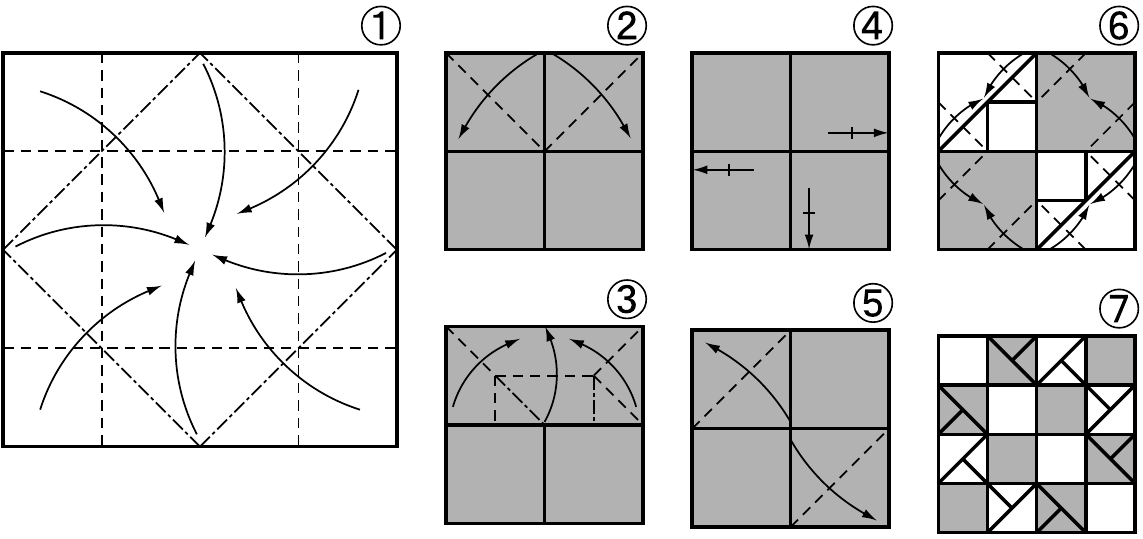}
\begin{itemize}
\item Step 1: collapse with preliminary folds.
\item Step 2: open one layer.
\item Step 3: fold back.
\item Step 4: repeat three times (care of the symmetry of the model).
\item Steps 5 and 6: bookfolds for color change.
\end{itemize}
  \caption{The optimal $4 \times 4$ static checkerboard, from
  a $8 \times 8$ square of paper.}
  \label{fig:s44}
\end{figure}

The optimal case here corresponds to a $8 \times 8$ square paper.
This is the first checkerboard that allows to avoid edge effects
while being optimal.
A compact folding sequence of the
original design of Max Hulme \cite{Hulme1985}
is depicted in figure~\ref{fig:s44}.

\clearpage
\subsection{$5 \times 5$ checkered board}

\begin{figure}[htbp]
  \centering
  \includegraphics[scale=1]{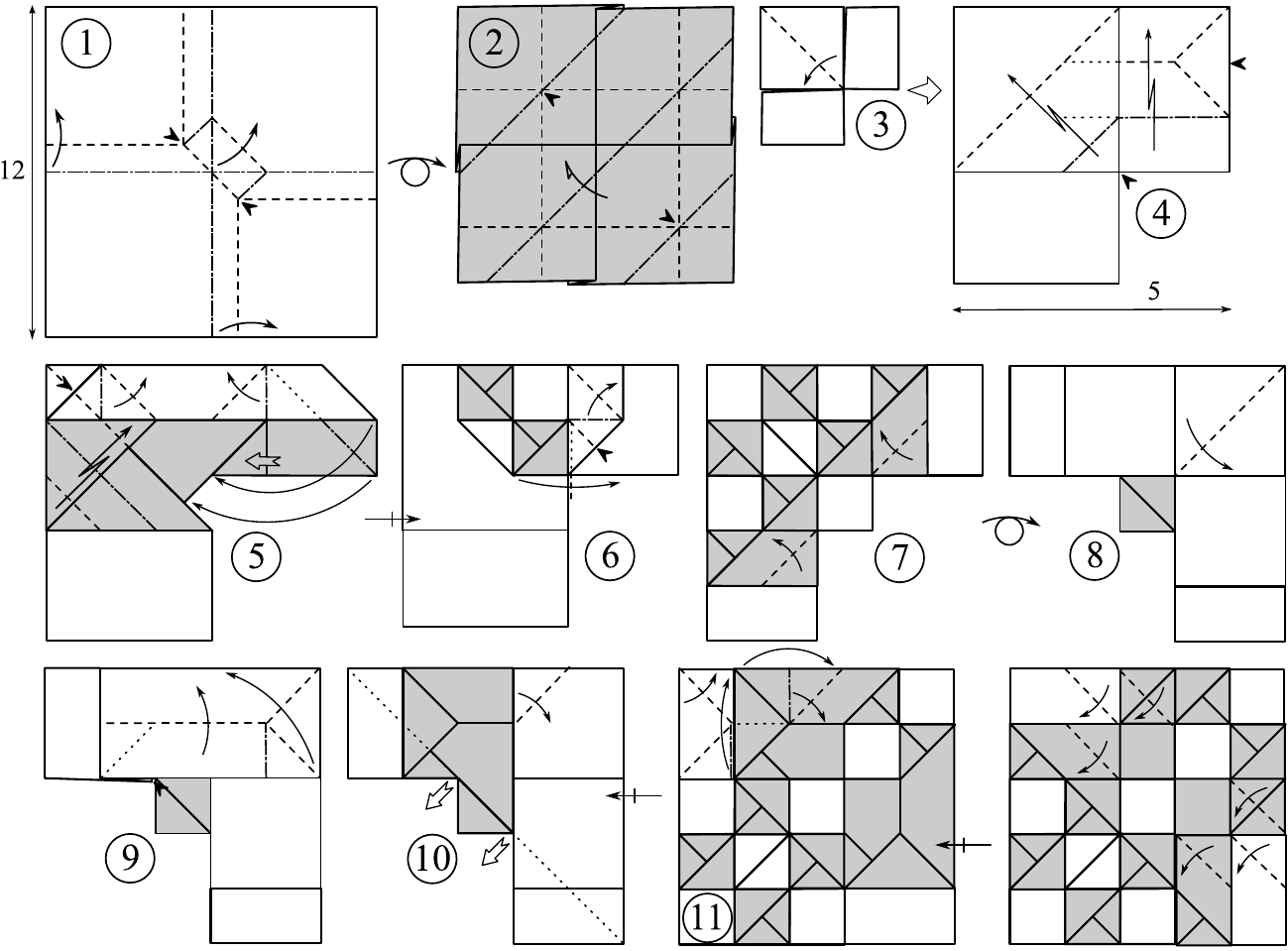}
\begin{itemize}
\item Step 1: fold the $12 \times 12$ paper. Turn the model.
\item Step 2: distorted preliminary-like base.
\item Step 3: fold the flap.
\item Step 4: 2 coupled squashes.
\item Step 5: squash and crimp on the left; swivel on the right.
\item Step 6: swivel; repeat steps 3-6.
\item Step 7: flip for color change. Turn the model.
\item Step 8: fold the flap.
\item Step 9: fold.
\item Step 10: flip for color change; repeat steps 8-10. 
  Unfold the back face along main diagonal.
\item Step 11: 2 couples swivels.
\item Step 12: flip for color change.
\end{itemize}
  \caption{The optimal $5 \times 5$ static checkerboard, from
  a $12 \times 12$ square of paper.}
  \label{fig:s55}
\end{figure}

The optimal case corresponds to a $12 \times 12$ square paper.
figure~\ref{fig:s55} proposes a solution.


\begin{figure}[htbp]
  \centering
  \includegraphics[scale=1]{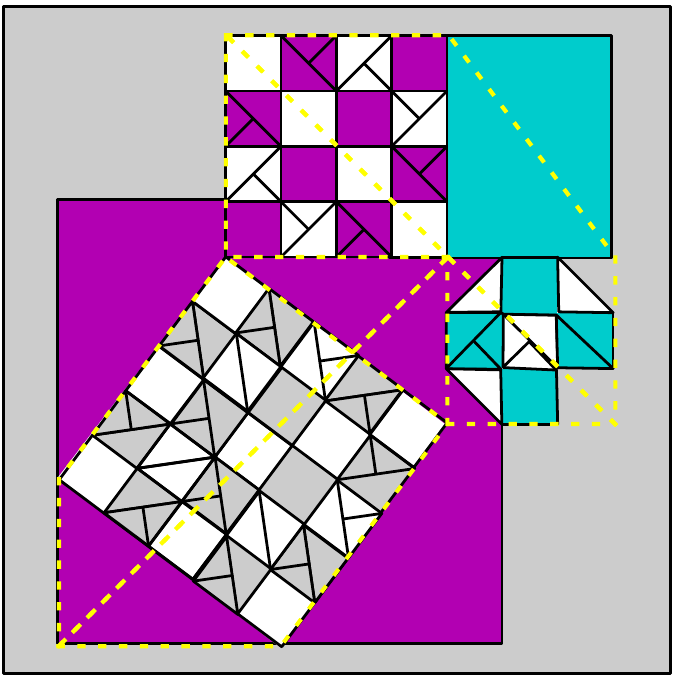}
  \caption{The Pythagorean theorem with its proof-without-word of Da Vinci.}
  \label{fig:pythagore}
\end{figure}

Finally, a geometric picture 
(with the checker edges $3^2 + 4^2 = 5^2$ and 
their optimal square edges $4 + 8 = 12$)
allows to illustrate the Pythagorean theorem, together with
the puzzle-like proof of Da Vinci,
figure~\ref{fig:pythagore}.

\clearpage
\subsection{$8 \times 8$ chessboard}

\begin{figure}[htbp]
  \centering
  \includegraphics[scale=1]{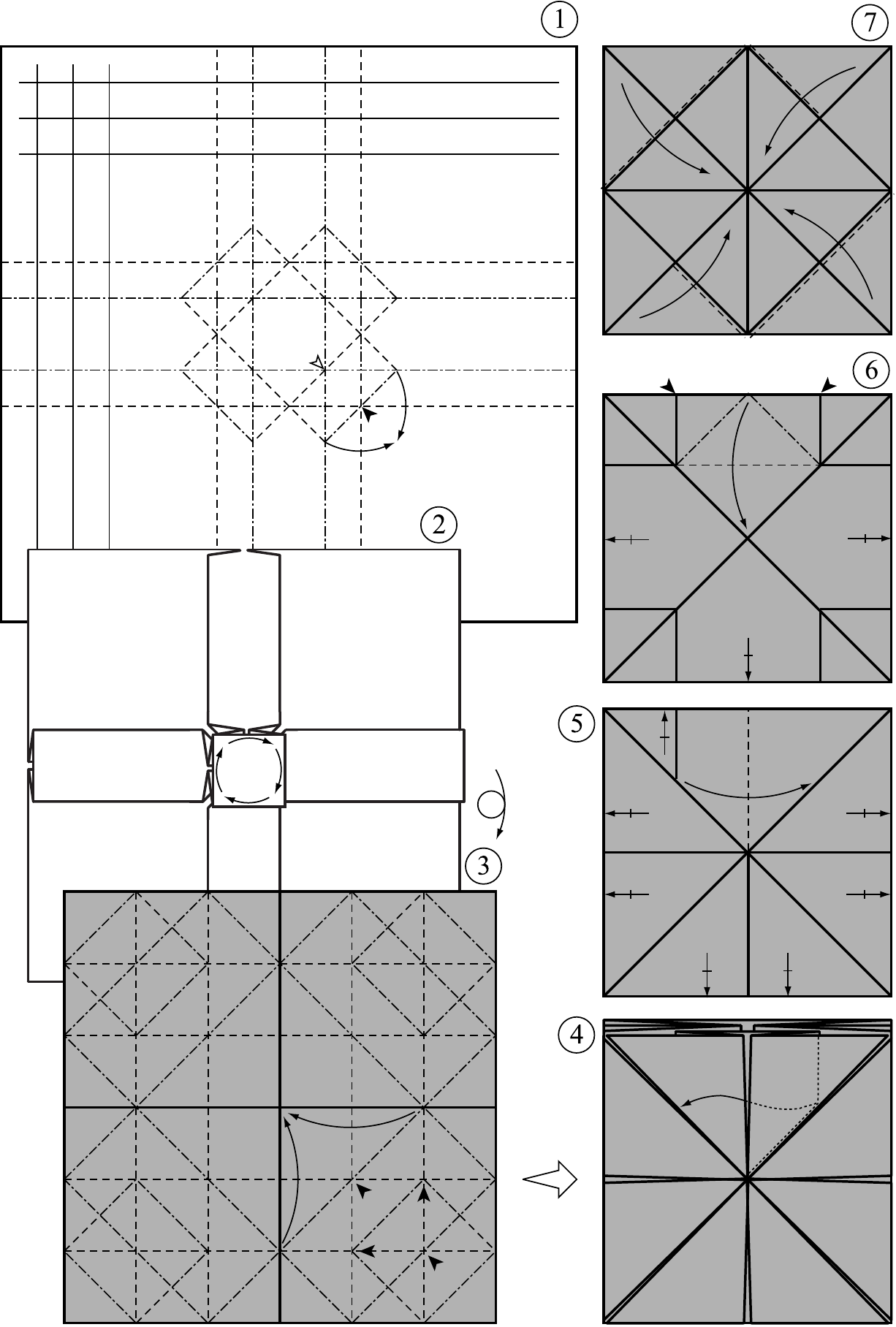}
  \caption{The optimal $8 \times 8$ static chessboard, from
  a $32 \times 32$ square of paper.}
  \label{fig:s88a}
\end{figure}

\begin{figure}[htbp]
  \centering
  \includegraphics[scale=1]{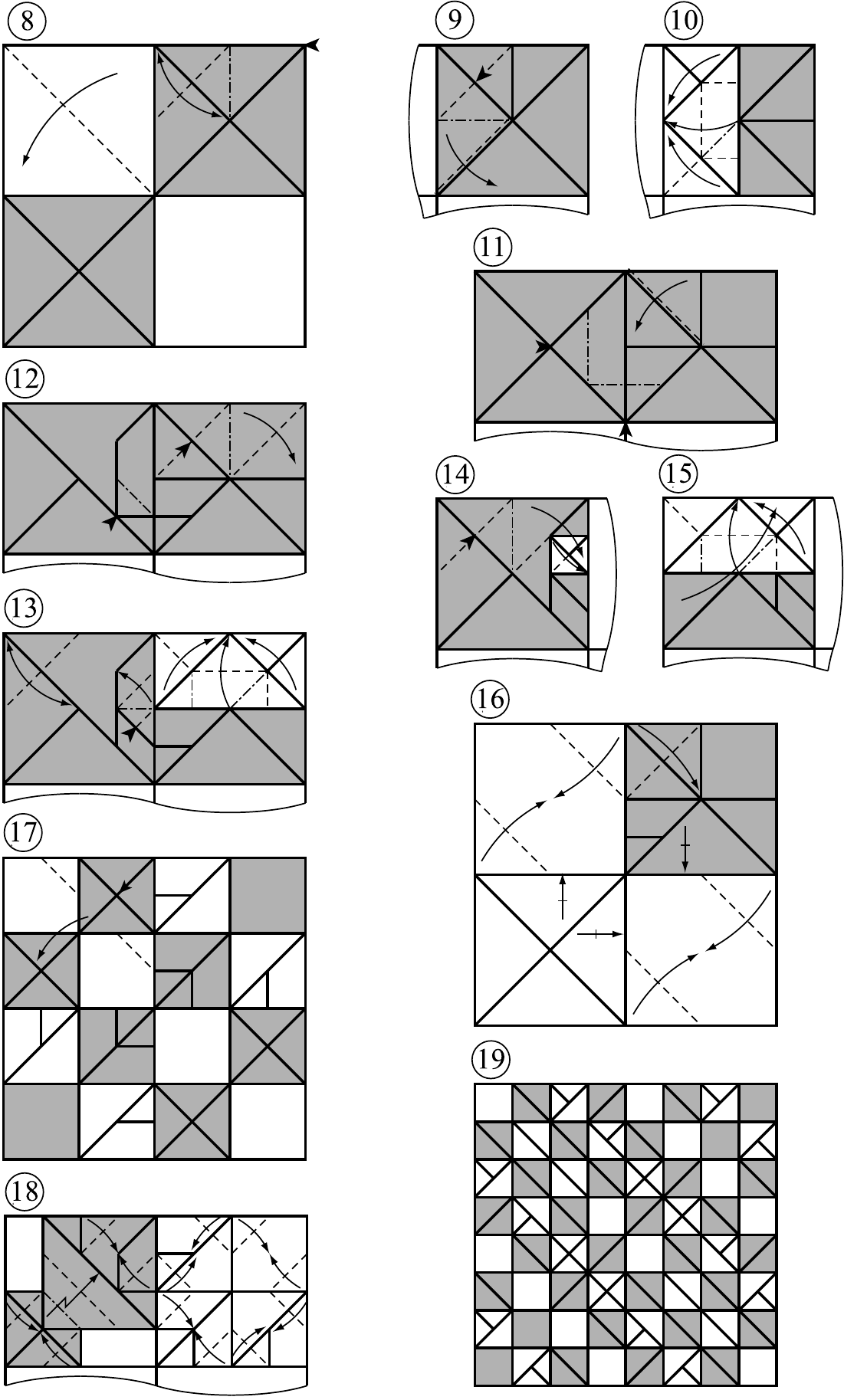}
  \caption{The optimal $8 \times 8$ static chessboard, from
  a $32 \times 32$ square of paper. Continued.}
  \label{fig:s88b}
\end{figure}

The following model has been published in \cite{Dureisseix2000}
and produces a $8 \times 8$ chessboard from a 
$32 \times 32$ square of paper. It is therefore expected to be
optimal.
The folding sequence of figures~\ref{fig:s88a} and \ref{fig:s88b} is:
\begin{itemize}
  \item Step 1: pre-crease the 1/16th. Waterbomb-base folds.
  \item Step 2: rotate to lock. Turn over.
  \item Step 3: bring edges again toward center.
  \item Step 4: sink.
  \item Step 5: book fold. Repeat steps 4-5.
  \item Step 6: petal fold. Repeat.
  \item Step 7-8: fold flaps. Here is a $2 \times 2$ pattern.
  \item Step 9: squash and swivel together.
  \item Step 10: petal fold (note the dissymmetry of the fold).
  \item Step 11: open sink on the left. Reverse fold on the bottom.
  \item Step 12: open sink again on the left. Same fold as in step 9 on the right.
  \item Step 13: squash-swivel on the left. Same fold as in step 10 on the right.
  \item Step 14: same fold as in step 9.
  \item Step 15: same fold as in step 10.
  \item Step 16: color change. Repeat steps 8-16 according to symmetries.
  \item Step 17: here is now a $4 \times 4$ pattern. Re-fold wider.
  \item Step 18: color change. Repeat steps 17-18 everywhere needed.
\end{itemize}

\clearpage
\subsection{Generic construction of an $n \times n$ checkerboard}

\begin{figure}[htbp]
  \centering
  \includegraphics[scale=2]{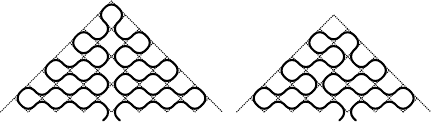}
  \caption{Edge diagrams for $n=8$ and $n=7$
  (half top parts only).}
  \label{fig:graphnncheckerboard}
\end{figure}

A generic folding sequence can be built for a general 
$n \times n$ checkerboard.
The design is not the most elegant, but it can be extended
to any number $n \ge 5$.
The folding sequence relies of a strip circulating on the
folded model than leads to a `dendritic-like' edge pattern, 
depicted in figure~\ref{fig:graphnncheckerboard}
for $n=8$ and $n=7$ cases.

\begin{figure}[htbp]
  \centering
  \includegraphics[scale=1]{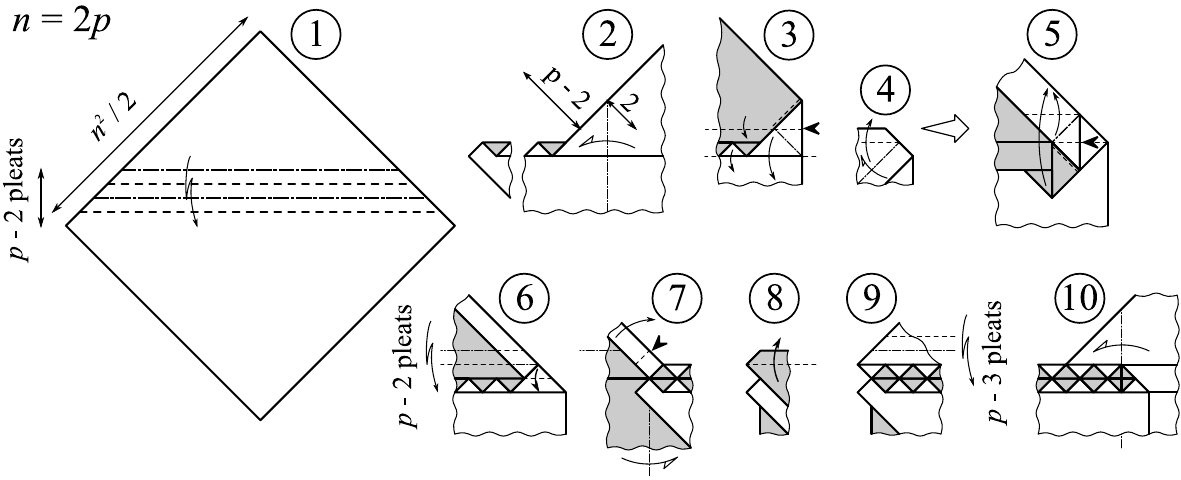}
\begin{itemize}
  \item Step 1: 
  after the pre-crease of a $n \times n$ grid, with all diagonals 
  pre-creased as well,
  make $p-2$ pleats along the diagonal direction.
  \item Step 2: book fold back.
  \item Step 3: squash and book folds.
  \item Step 4: swivel fold.
  \item Step 5: a complex squash and swivel coupled fold.
  \item Step 6: flip on the right and make $p-2$ inside reverse
  folds to make the same number of pleats.
  \item Step 7: outside reverse fold.
  \item Step 8: book fold.
  \item Step 9: make $p-3$ pleats.
  \item Step 10: book fold back (this step is similar to step 2).
  Repeat by recursion from step 3, replacing $p$ by $p-1$.
\end{itemize}
  \caption{The generic folding sub-sequence for the optimal
  $n \times n$ static checkerbord, for $n$ even
  (only one fourth of the square is depicted due to symmetries).}
    \label{fig:modelnncheckerboardEven}
\end{figure}

The $n=2p$ even case has two symmetries with respect to the diagonals 
of the board. 
The sequence is the repetition of a generic sub-sequence,
which is described in
figure~\ref{fig:modelnncheckerboardEven}.

\begin{figure}[htbp]
  \centering
  \includegraphics[scale=1]{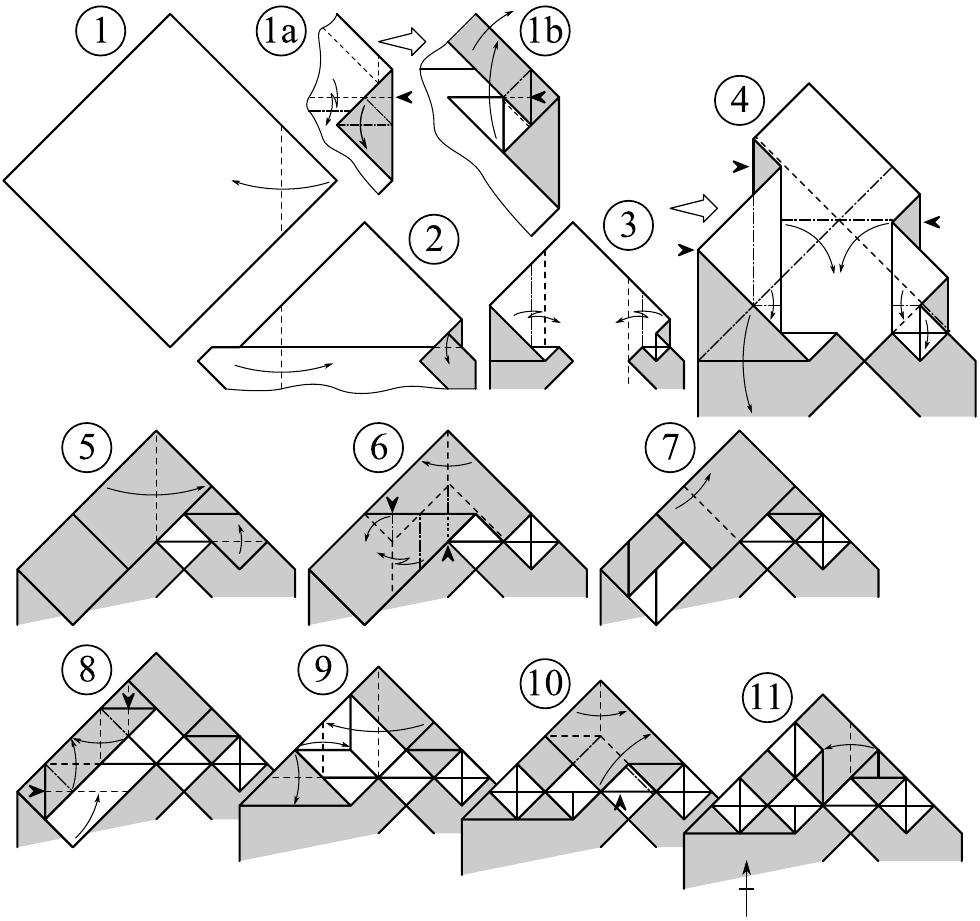}
\begin{itemize}
  \item Step 1: book fold one-third of the $12 \times 12$ square.
  \item Step 1a: make a pleat and a squash fold. 
  \item Step 1b: squash and unfold.
  \item Step 2: book fold, and flip.
  \item Step 3: make two underlying pleats.
  \item Step 4: preliminary fold, and two swivel folds.
  \item Step 5: half fold and flip.
  \item Step 6: fold back with a squash and a pleat.
  \item Step 7: book fold.
  \item Step 8: two swivel folds.
  \item Step 9: flips.
  \item Step 10: inside reverse fold, and book fold back.
  \item Step 11: flip and repeat on the bottom 
  for the full $5 \times 5$ design.
\end{itemize}
  \caption{Another folding sequence of the optimal $5 \times 5$ 
  static chessboard, from   a $12 \times 12$ square of paper.}
  \label{fig:modelnncheckerboardLast}
\end{figure}

\begin{figure}[htbp]
  \centering
  \includegraphics[scale=1]{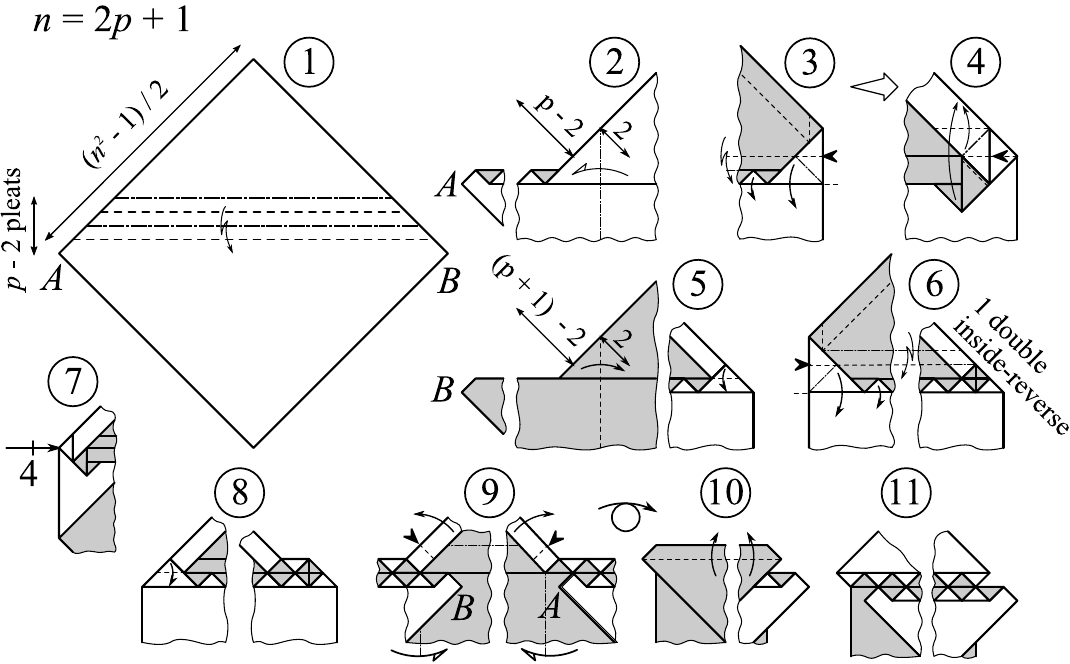}
\begin{itemize}
  \item Step 1: 
  after the pre-crease of a $n \times n$ grid, with all diagonals 
  pre-creased as well,
  make $p-2$ pleats along the diagonal direction.
  \item Step 2: book fold back.
  \item Step 3: squash and book fold, with one pleat at the back.
  \item Step 4: a complex squash and swivel coupled fold.
  \item Step 5: flip on the right and book fold back on the left.
  \item Step 6: one double inside reverse fold on the right to make
  a pleat, and repeat step 3 on the left.
  \item Step 7: repeat step 4 on the left.
  \item Step 8: flip on the left.
  \item Step 9: outside reverse folds. Return.
  \item Step 10: book fold.
  \item Step 11: the obtained result is similar to step 1.
  The entire folding sub-sequence 1-11 can now be repeated recursively
  (replacing $p$ by $p-1$, and switching points $A$ and $B$),
  until step 3 of figure~\ref{fig:modelnncheckerboardLast}
  is obtained, leading to the last stages of this figure.
\end{itemize}
  \caption{The generic folding sub-sequence for the optimal
  $n \times n$ static checkerbord, for $n$ odd
  (only one half of the square is depicted due to symmetry).}
    \label{fig:modelnncheckerboardOdd}
\end{figure}

The odd case is more difficult than the even case since it 
possess less symmetries, and is described herein.
The sequence is also the repetition of a generic sub-sequence,
except for the last stage that leads to complete a 
$n=5$ half checkerboard which is of
particular design as a kind of edge effect.
This special $5 \times 5$ is depicted in 
figure~\ref{fig:modelnncheckerboardLast}, 
and the constructive proof in
figure~\ref{fig:modelnncheckerboardOdd}.

\begin{figure}[htbp]
  \centering
  \includegraphics[scale=0.5]{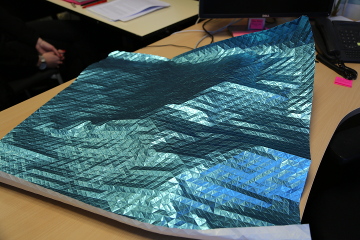}
  \includegraphics[scale=0.45]{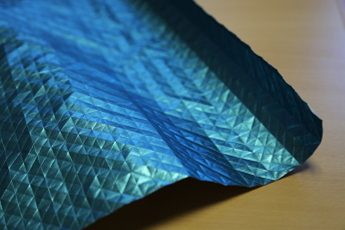}
  \caption{A $64 \times 64$ pre-creased square of paper, 
  used for the $11 \times 11$ checkerboard design. 
  Photo by INSA Lyon, 2017.}
      \label{fig:prepare11}
\end{figure}

\begin{figure}[htbp]
  \centering
  \includegraphics[scale=0.65]{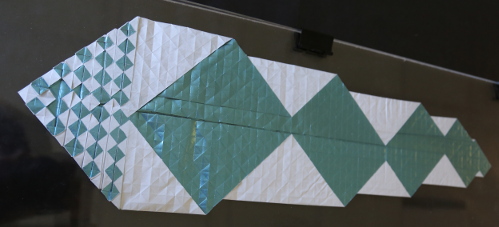}
  \caption{Folded $11 \times 11$ checkerboard from a
  single $60 \times 60$ square of paper, one color on each face
  (half model, second half remaining to be folded). 
  Photo by INSA Lyon, 2017.}
      \label{fig:final11}
\end{figure}

Following this process, a $11 \times 11$ checkerboard has been
folded that is probably the largest optimal checkerboard
that has been effectively folded.
It is photographed in figures~\ref{fig:prepare11} 
and \ref{fig:final11}. 

\end{document}